\documentclass[12pt]{iopart}
\usepackage{iopams}
\usepackage{epsf}

\usepackage{graphicx}% Include figure files
\usepackage{dcolumn}% Align table columns on decimal point
\usepackage{bbm}% bold math
\usepackage{multirow}

\def\Ot{\tilde\Omega t}

\def\x1{\frac{1}{\xi} }
\def\xt{\xi t}
\def\Om{\tilde\Omega}
\def\cs{\cos \sigma t}
\def\sz{\sin \sigma t}
\def\zr{\cos \rho t}
\def\sr{\sin \rho t}

\begin{document}

\title{One pendulum to run them all}

\author{G. Barenboim\dag\  and J.A. Oteo\ddag\ }

\address{\dag Departament de F\'{\i}sica Te\`{o}rica  and Instituto de
F\'{\i}sica Corpuscular, Universitat de Val\`{e}ncia,
46100-Burjassot, Val\`{e}ncia, Spain}

\address{\ddag\ Departament de F\'{\i}sica Te\`{o}rica, Universitat de
Val\`{e}ncia,  46100-Burjassot, Val\`{e}ncia, Spain}

\ead{gabriela.barenboim@uv.es, oteo@uv.es}

\begin{abstract}
{The analytical solution of the three--dimensional linear pendulum in a rotating frame of reference is obtained,
including Coriolis and centrifugal accelerations, and expressed in terms of initial conditions. This result offers
the possibility of treating Foucault and Bravais pendula as trajectories of the very same system of
equations, each of them with particular initial conditions. We compare with the common two--dimensional approximations
in textbooks. A previously unnoticed pattern in the three--dimensional
Foucault pendulum attractor is presented.  }
\end{abstract}

\pacs {}




\maketitle

\section{Introduction\label{sec:intro}}

During the first year(s) of undergraduate studies students are constantly warned to work
in an inertial frame, as only in this case Newton's second Law can be safely applied.
Since it rotates around its axes (even ignoring its motion around the sun), a reference
frame fixed to the Earth is definitively non inertial. However, for most mechanics-related
results this fact can be overlooked, as this rotation is not fast.
But there are certain circumstances when non inertial reference frames, specially rotating
frames, look like an interesting thing to use. This is specially so, when trying  to study movement
on Earth surface or when looking to prove
or measure Earth motion itself through a mechanical experiment.

The first  proof of Earth's rotation, and perhaps the most simple and elegant,
is indeed the Foucault pendulum experiment \cite{Foucault} performed in 1851.
A handwriting annotation by Foucault founded among his papers reads: \emph{Mercredi, 8 janvier, 2 heures du matin: le pendule a tourn\'e dans le sens du mouvement diurne de la sph\`ere c\'eleste} \cite{travaux}. Namely, the experiment succeeded on January 8th at 2 hours \emph{a.m.}
It was carried out in his home's cellar, d'Assas street in Paris, using a 2 m. long pendulum \cite{Tobin}. Afterwards, in February, the experiment was repeated at a larger scale with a 11 m. long pendulum in the Paris Observatory \cite{Foucault} under the permission of Arago. It is worth noting the deep acknowledgement Foucault  expresses toward his assistant Froment.

Nowadays it is widely known the public exhibition of the pendulum held in March, on the request and backing of the prince--president of the Republic.  Gracefully suspended from the
ceiling of the dome of the Panth\'eon in Paris,
it consisted of a 28 kg brass-coated lead bob attached to a 67 meter long steel aircraft cable. The plane of the pendulum's swing appeared to rotate clockwise 264 degrees every day (11 degrees per hour), sweeping a circle of 16 m. diameter.
Since no rotational forces act on the pendulum, it is obvious that it is the Earth beneath that is actually rotating.
However sizeable effects, at naked eye, take several hours to become apparent. We will go back to this point later.

On the very same year (and amazingly also in Paris) Bravais proposed an alternative
way to observe the rotation of the Earth, using also pendula \cite{Bravais1, Bravais}. His approach was
however different and has been review recently in this journal \cite{BM}. He explored the difference
in the period of oscillations of right and left handed pendula. This fact, as we are going
to show later, opens the door to
observe the rotation of the Earth in a few minutes.

Yet, despite not receiving a unified treatment, Foucault and Bravais pendula share the same
equations of motion, differing only through  initial conditions.
Therefore they may be considered, in all senses, as
two particular cases of a large variety of oscillations modes of a
pendulum in a non inertial reference frame.

Foucault type oscillations are obtained when the bob is moved away
from the equilibrium position
and then released with zero initial velocity. In an inertial reference
frame the pendulum would oscillate
in a fixed plane. Due to the Earth rotation, we observe this plane to
rotate. Thus the trajectory followed
by the bob produces the well-known beautiful patterns,
which are the proof of Earth's rotation.
Instead, Bravais pendulum refers to conical oscillations of the bob.
To this end, the bob is released from a non equilibrium point with a
precise tangential speed in such a way that the projection of the trajectory
on the horizontal plane describes simply a circumference.
What renders  Bravais pendulum interesting is that the
trajectories are not invariant under sign reversal in the
initial tangential velocities, as a consequence of Earth's rotation.
It is just this phenomenon that led Bravais to propose the pendulum
as an experimental demonstration of Earth's rotation.

Being just two particular sets of initial condition for the same system of
equations, it becomes obvious that there is a whole plethora of systems
which presents interesting features worth looking at. Specially knowing
that extremely precise initial conditions can be quite challenging to achieve from an
experimental point of view.

One of the purposes of this paper is to give an explicit solution of the
linear pendulum in terms of initial conditions. We find it of particular interest
for students whose training in solving linear systems of differential equations
is concurrent with their course in mechanics.

The structure of the paper is the following. In Section \ref{sec:ecs}
we collect the equations of motion. In Section \ref{sec:2D} we review some common
approximate solution presented in textbooks. Next, in Section \ref{sec:exact} the analytical
exact solution in three dimensions is obtained. Section \ref{sec:Chevilliet} connects our results with a nowadays almost forgotten theorem by Chevilliet.
In Section \ref{sec:2D3D} a comparison two-- and three--dimensional treatment of the Foucault pendulum is carried out. A previously unnoticed pattern in the attractor is reported. Eventually, Section \ref{sec:fin} contains a discussion of our results.

\begin{figure}[t]
\centering
%\begin{center}
\includegraphics[scale=0.6]{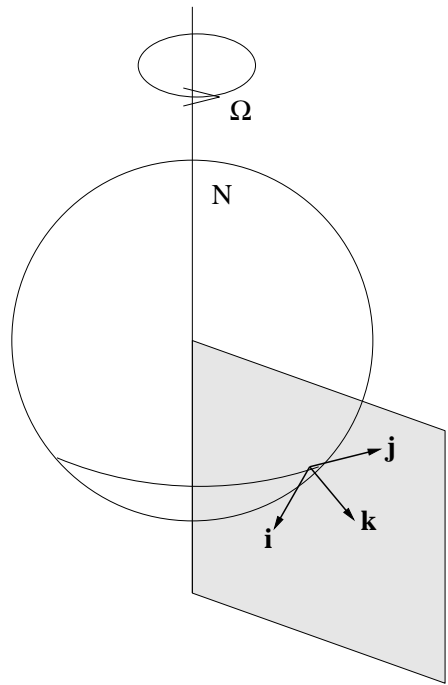}
\caption{Reference system attached to the rotating Earth used to study the pendulum.}
\label{fig:fig1}
%\end{center}
\end{figure}

\begin{figure}[t]
\centering
%\begin{center}
\includegraphics[scale=0.6]{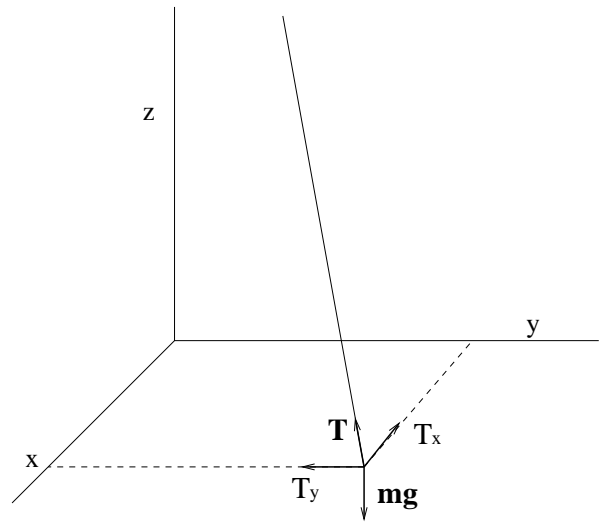}
\caption{For a simple pendulum with small amplitude approximations gives $T_i = - T\; i/l $, with $i=x,y,l-z$ and $T\simeq mg$.}
\label{fig:fig2}
%\end{center}
\end{figure}

\section{Equations of motion of the pendulum for small amplitudes}\label{sec:ecs}

We will follow the careful exposition in \cite{BB} to obtain the equations of motion of a pendulum with small oscillations, referred to a rotating reference frame. The starting point is the Newton equation for the bob of the pendulum
\begin{equation}\label{eq:Newton}
    m\ddot{\vec r}=\vec{T}-mg\hat{k},
\end{equation}
where the dot stands for time derivative,  $\vec r$ is the position vector of the bob of mass $m$ with respect to an inertial system of reference,
and $\vec T$ is the tension of the wire. In this instance,  $\hat k$ is the unit vector directed vertically upward, which follows the radial direction of the Earth, assumed to have spherical symmetry. This remark will be of interest momentarily. Eventually, $g$ is the corresponding intensity of  gravity.

The analysis of Foucault pendulum is carried out in a rotating reference system fixed to the Earth surface. We will consider that the Earth turns with constant angular velocity
\begin{equation}\label{eq:Earth}
    \vec{\Omega}=-\hat{\imath} \Omega \cos \beta +\hat{k}\Omega \sin \beta,
\end{equation}
in terms of the latitude $\beta$, and $\Omega=2\pi/86400 \; {\rm rd/s}  =0.000073 \; {\rm rd/s}$. The unit vector $\hat \imath$ is tangent to the local  meridian and points toward the equator, in the northern hemisphere. The third unit vector, $\hat \jmath$, is directed eastward (see Figure \ref{fig:fig1}).

In a rotating reference system with origin located at the center of the Earth, time derivatives have to be replaced with more involved expressions. In particular, if $\dot{\vec\Omega}=0$, the acceleration $\ddot{\vec r}$ becomes $\ddot{\vec r}+2\vec{\Omega}\times \dot{\vec r}+{\vec \Omega}\times({\vec \Omega}\times{\vec r})$. The term $2\vec{\Omega}\times \dot{\vec r}$ is the Coriolis acceleration, and $\vec{\Omega}\times({\vec \Omega}\times{\vec r})$ stands for the centrifugal acceleration. Notice that both contributions are first and second order in $\Omega$, respectively; and since $\Omega \ll 1$, they can be considered small corrections.

If we move to a frame of reference fixed on the surface of the Earth at position $\vec R$, then we have to replace
$\vec{r}$ with $\vec{r}+\vec R$, in the transformation above. The equation of motion in the rotating system reads then
\begin{equation}\label{eq:Newtonrot}
     m\ddot{\vec r}=\vec{T}-mg\hat{k}-m{\vec \Omega}\times({\vec \Omega}\times{\vec R)}-
     2m\vec{\Omega}\times \dot{\vec r}-m{\vec \Omega}\times({\vec \Omega}\times{\vec r}) ,
\end{equation}
where, once again, $\vec r$ is now referred to the frame of reference on the surface of the Earth.
The term ${\vec g}_{e}\equiv g\hat{k}+m{\vec \Omega}\times({\vec \Omega}\times{\vec R)} $
defines the so--called \emph{effective gravity}, which takes into account the slight effect of centrifugal acceleration (which is order $\Omega^2$).
The vector ${\vec g}_{e}$ determines the local \emph{plumb line} and allows us to simplify the structure of (\ref{eq:Newtonrot}) by redefining the unit vector $\hat k$ so as to follow  the direction of
${\vec g}_{e}$. This way, ${\vec g}_{e}$ has only one non--vanishing component (its value varies slightly with latitude)
and is everywhere orthogonal to the surface of the Earth.
We are implicitly stating that the Earth is not a sphere but its real shape is given by the equipotential surface resulting from the interplay between the gravitational and centrifugal interactions, the geoid.
The definition of the unit vectors $\hat \imath,\hat \jmath$ does not change. They are, of course, orthogonal to $\hat k$.
These considerations simplify (\ref{eq:Newtonrot}) to the form
\begin{equation}\label{eq:Newtonrotfin}
     m\ddot{\vec r}=\vec{T}-mg\hat{k}-
     2m\vec{\Omega}\times \dot{\vec r}-m{\vec \Omega}\times({\vec \Omega}\times{\vec r}),
\end{equation}

The linear approximation for the pendulum equations corresponds to assume that the tension vector is given by (see Figure \ref{fig:fig2})
\begin{equation}\label{eq:tension}
    {\vec T}= -\frac{mg}{l}[x\hat i+y \hat j +(l-z)\hat k],
\end{equation}
valid for small amplitudes. Here, $l$ is the length of the pendulum, and the origin of coordinates is at its supporting point.  Introducing the notation $\omega\equiv \sqrt{g/l}$, we get  from (\ref{eq:Newtonrotfin}) and (\ref{eq:tension})
\begin{equation}\label{eq:Nvec}
     \ddot{\vec r}=\omega^2\vec{r}-
     2\vec{\Omega}\times \dot{\vec r}-{\vec \Omega}\times({\vec \Omega}\times{\vec r}),
\end{equation}
or, explicitly in components
\begin{eqnarray} \label{eq:F3}
\ddot x&=&-\omega^2x+2s\Omega\dot y+s^2\Omega^2x+sc\Omega^2z, \nonumber \\
\ddot y&=&-\omega^2 y-2s\Omega\dot x -2c\Omega \dot z +\Omega^2 y ,\\
\ddot z&=& -\omega^2 z+2c\Omega\dot y+cs\Omega^2x+c^2\Omega^2z ,\nonumber
\end{eqnarray}
where we have used the notation
\begin{equation}\label{eq:sc}
    s\equiv \sin \beta, \qquad c\equiv \cos \beta .
\end{equation}
We remind the reader that in these equations the $z$ axis is vertical upward, the $x$ axis is north-to-south, and
the $y$ axis is west-to-east.

The approximation in (\ref{eq:tension}) is crucial from a practical point of view, for it leads to
the linear differential system (\ref{eq:F3}), which is homogeneous with constant coefficients, and therefore
solvable.
One standard solution method consists in transforming (\ref{eq:F3})
into a linear differential equation of sixth order. Else, one can alternatively transform
(\ref{eq:F3}) into a system of first
order and dimension six. Both procedures are indeed cumbersome and,
to the best of our knowledge,
no explicit exact solution is found in textbooks.

\section{Approximate trajectories in two dimensions} \label{sec:2D}

In addition to the assumption of
small amplitude oscillations implicit in (\ref{eq:tension}),
two further approximations are usually carried out in order to obtain approximate trajectories.
Firstly, one can assume $\Omega \ll \omega$, and then
ignore the centrifugal acceleration, namely the quadratic terms $\Omega^2$ in (\ref{eq:F3}).
This provides a remarkable simplification.
A further approximation considers
that the pendulum length $l$ is large enough so as the motion of the pendulum takes place
in a plane. Thus $z,\dot z,\ddot z$ are dropped in (\ref{eq:F3}) and the equations of
motion of the planar pendulum in the small amplitude approximation (without centrifugal acceleration) reduce to the two--dimensional system
\begin{eqnarray} \label{eq:F2}
\ddot x&=&-\omega^2x+2s\Omega\dot y , \nonumber \\
\ddot y&=&-\omega^2 y-2s\Omega\dot x .
\end{eqnarray}
Due to the symmetry of these equations a convenient way to solve them is to introduce
the complex representation $Z=x+iy$. Then the system (\ref{eq:F2})
may be written down in terms of a sole complex equation
\begin{equation}
\dot Z+2i\Om\dot Z+\omega^2Z=0,
\end{equation}
with $\Om \equiv s\Omega$. This is a linear homogeneous differential equation of second degree whose
solution reads $Z(t)=A\exp(\lambda_+ t)+B\exp(\lambda_- t)$; and $A,B$ are complex
arbitrary constants. Here $\lambda_\pm\equiv -i(\Om \pm\xi), \xi\equiv \sqrt{\omega^2+\Om^2}$,
are the (complex) roots of the second degree equation
$\lambda^2+2i\Om \lambda +\omega^2=0$. The complex solution becomes
\begin{equation} \label{eq:Z}
Z(t)=\exp(-i\Om t)[A\exp(i\xi t)+B\exp(-i\xi t)].
\end{equation}
The four real quantities contained in the two complex constants $A,B$ are determined
so as $x(t), y(t)$, satisfy the given initial conditions $x(0),\dot x(0),y(0), \dot y(0)$.
This requires some further algebra, but just to interpret the
structure of the solutions let us think of initial conditions such that $A$ and $B$
are complex conjugate of each other.
In that case the terms in square brackets stand for a real function and the solutions $x(t),y(t)$
may be simply written down as
\begin{eqnarray}
x(t)&=&a\cos(\Om t)\cos(\omega t+\phi), \nonumber\\
y(t)&=&a\sin(\Om t)\cos(\omega t+\phi),
\end{eqnarray}
in terms of an amplitude $a$, a phase $\phi$, and where, in addition,
we have approximated $\xi\simeq \omega$. Thus the trajectories as a function of $t$ are the product
of a fast oscillation of frequency $\omega$ describing the natural motion of a simple pendulum,
and a slow oscillation of frequency $\Om$ which originates a rotation of the pendulum oscillation plane.

By using elementary
matrix methods, the general solution of (\ref{eq:F2}) may be expressed in terms of
initial conditions. To get rid of the complex representation in (\ref{eq:Z})
we write $A\equiv a_r+ia_i, B\equiv b_r+b_i$, in terms of new four real parameters $a_r,a_i,b_r,b_i$.
Next, using Euler formula in $Z(t)$ and $\dot Z(t)$,  equation (\ref{eq:Z}) yields  after some algebra
\begin{equation} \label{eq:Mt}
\fl
\left(
  \begin{array}{c}
    x(t) \\
    \dot x(t) \\
    y(t) \\
    \dot y(t) \\
  \end{array}
\right)
=
\left(
  \begin{array}{cccc}
    \cs & -\sz & \zr & \sr \\
    -\sigma\sz & -\rho\cs & -\rho\sr & \rho\zr \\
    \sz & \cs & -\sr & \zr \\
    \sigma\cs & -\sigma\sz & -\rho\zr & -\rho\sr \\
  \end{array}
\right)
\left(
  \begin{array}{c}
    a_r \\
    a_i \\
    b_r \\
    b_i \\
  \end{array}
\right)
\equiv M(t)
\left(
  \begin{array}{c}
    a_r \\
    a_i \\
    b_r \\
    b_i \\
  \end{array}
\right)
\end{equation}
where $\sigma \equiv \xi-\Om , \rho\equiv \xi+\Om$,
with $\xi$ already defined above. Equation (\ref{eq:Mt}) defines the time dependent
matrix $M(t)$.
At time $t=0$ we have
\begin{equation}
\left(
  \begin{array}{c}
    x(0) \\
    \dot x(0) \\
    y(0) \\
    \dot y(0) \\
  \end{array}
\right)
=M(0)
\left(
  \begin{array}{c}
    a_r \\
    a_i \\
    b_r \\
    b_i \\
  \end{array}
\right)
\end{equation}
Therefore, we can now give the trajectories in terms of initial conditions
\begin{equation}
\left(
  \begin{array}{c}
    x(t) \\
    \dot x(t) \\
    y(t) \\
    \dot y(t) \\
  \end{array}
\right)
=M(t) M^{-1}(0)
\left(
  \begin{array}{c}
    x(0) \\
    \dot x(0) \\
    y(0) \\
    \dot y(0) \\
  \end{array}
\right) .
\end{equation}
A straightforward computation yields
\begin{equation}
M^{-1}(0)=\left(%
\begin{array}{cccc}
  \rho & 0 & 0 & 1 \\
  0 & -1 & \rho & 0 \\
  \sigma & 0 & 0 & -1 \\
  0 & 1 & \sigma & 0 \\
\end{array}%
\right) .
\end{equation}
The explicit solution for the planar, small amplitude pendulum, with Coriolis and without
centrifugal accelerations reads then
\begin{eqnarray} \label{eq:traj-x}
x(t)= &\fl&  \nonumber
\frac{1}{\xi}[(\xi\cos \Ot \cos \xt + \Om \sin \Ot \sin \xt )x_0+
(\cos \Ot \sin \xt ) \dot x_0+ \\ &&
(\xi\sin\Ot \cos \xt - \Om \cos\Ot \sin \xt) y_0+  (\sin\Ot \sin\xt) \dot y_0],\\
y(t)=&\fl&\nonumber  \label{eq:traj-y}
-\frac{1}{\xi}[(\xi\sin \Ot\cos \xt +\Om\cos \Ot\sin\xt ) x_0 -
(\sin\Ot\sin\xt)\dot x_0+\\ &&
(\xi\cos\Ot \cos \xt + \Om \sin\Ot \sin \xt) y_0+  (\cos\Ot \sin\xt) \dot y_0],
\end{eqnarray}
where we use the notation $x(0)=x_0,y(0)=y_0,\dot x(0)=\dot x_0, \dot y(0)=\dot y_0$.

Equations (\ref{eq:traj-x}) and (\ref{eq:traj-y}) allow us to illustrate
the shape of trajectories for different sets of initial conditions, and in
particular those that give rise to the Foucault and Bravais pendula.
The Java applet freely downloadable at \cite{applet} allows an easy visualization of the trajectories for
arbitrary values of the parameters.

Next we illustrate the pendulum trajectories with three different choices of initial conditions. We have chosen the following parameter values: Length of the wire $L=100$ m., which gives the angular velocity $\omega =\sqrt{g/L}=0.3162$ rd/s; rotating reference system velocity $\Omega =\omega/10$;
and latitude $\beta =\pi /4$. The Earth rotation velocity is artificially speeded up for the sake of clarity in the figures.
In all the figures, the solid dot stands for the initial position. The arrow  establishes the direction of turn.

\subsection{Foucault pendulum in 2D}
The Foucault pendulum corresponds to the simplest possible choice of initial conditions. It is obtained
when we release the bob with vanishing initial velocity (in the Earth attached system) from, say, the point
$(x_0\ne 0,y_0=0)$. We illustrate in Figure \ref{fig:Fxy-1}, left panel, a trajectory in the plane $xy$.
The initial values are: $x_0=1, y_0=0, \dot x_0=\dot y_0=0$.
Also, in the right panel of Figure \ref{fig:Fxy-1} we plot the corresponding trajectory in the plane of velocities $\dot x \dot y$.

This case plainly and elegantly combines the motion of a single, planar pendulum with the precession of its
plane of oscillation. The deflexion of the plane of oscillation originates in the Coriolis acceleration.
Thus, the term $2\vec{\Omega}\times \dot{\vec{r}}$ deflects the trajectory to the right of the velocity vector
in the northern hemisphere. Hence,
every half oscillation of the pendulum produces an incremental effect.
The drawback of this set up is that, as we stated before, the
effects take several hours to become evident. For a Foucault pendulum located at the latitude
of Paris for example, it takes about 32 hours to complete a precession cycle.
In order to plot neat trajectory patterns we have used a much more greater angular velocity than the one corresponding to the value of Earth's rotation.

\subsection{Bravais pendulum in 2D}
The Bravais pendulum or conical pendulum, is doubtlessly the second simplest choice as regards to
the initial conditions. It is obtained when we release the bob from a point
$(x_0\ne 0, y_0=0)$ with an initial tangential velocity $(\dot x_0=0, \dot y_0=x_0(\omega-\tilde\Omega))$. Then the pendulum describes conical oscillations and the planar trajectory is a circumference.

If we reverse the initial velocity, $(\dot x_0=0, \dot y_0=-x_0(\omega-\tilde\Omega))$, the trajectory is only
conical-like. The planar projection is not a circumference any longer, but rather an open trajectory.
This is due to the effect of the rotation of the Earth on the motion of the pendulum bob.

This is illustrated in Figure \ref{fig:Bxy-1}. The black solid line corresponds to the conical pendulum released at $t=0$ with the initial conditions: $x_0=1,y_0=0, \dot x_0=0, \dot y_0=0.13142$. Whereas the red dotted line describes the trajectory started with reversed initial velocity.  In Figure \ref{fig:Bxy-1}, the left panel corresponds to the trajectory in the plane and in the right panel  we plot the corresponding trajectory in the plane of velocities $\dot x \dot y$.

Due to Earth's rotation, the
left handed and the right handed pendula  take slightly different times to complete one turn, besides the slight difference in the shape of their paths.
The delay is accumulative and their mutual phase difference becomes sizeable after
completing a moderated number of turns.
It therefore provides an easy and fast proof of Earth's rotation, at least
in an ideal experiment with two simultaneous pendula.

\subsection{General pendulum in 2D}
It is also interesting to observe the trajectory when the initial velocity has two non-vanishing components.
In Figure \ref{fig:Gxy-1} (left panel) we plot the trajectory corresponding to initial conditions $x_0=1,y_0=0,\dot x_0=-x_0(\omega-\Omega\sin \beta), \dot y_0=x_0(\omega-\Omega\sin \beta)$.
In the right panel of  Figure \ref{fig:Gxy-1} we represent the corresponding trajectory in the plane of velocities $\dot x \dot y$.

\begin{figure}[t]
\centering
%\begin{center}
\includegraphics[scale=0.8]{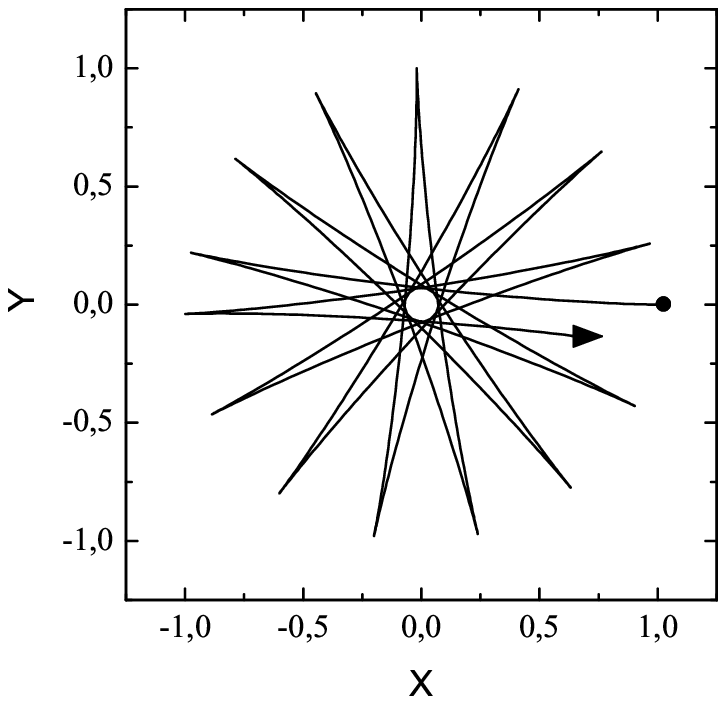}
\includegraphics[scale=0.8]{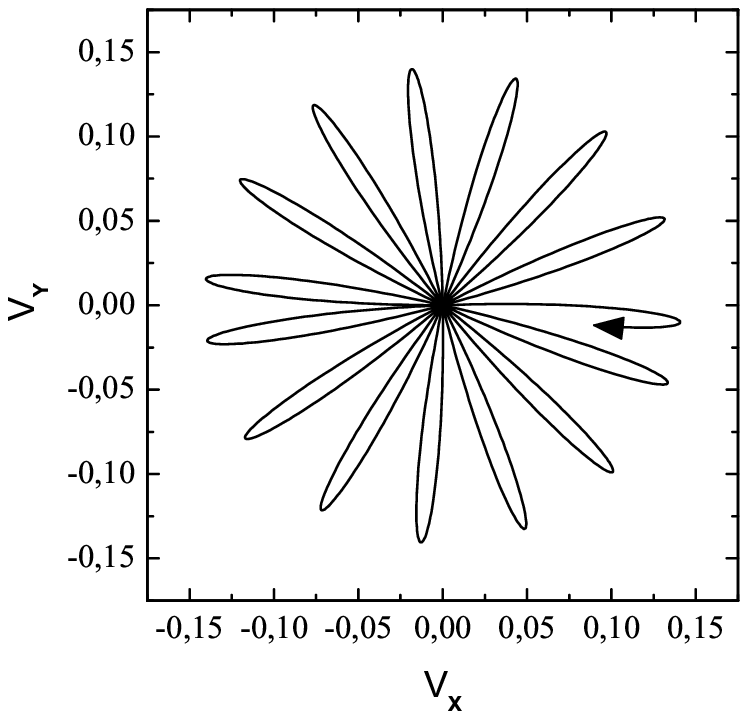}
\caption{Foucault pendulum in two dimensions. Trajectory in phase plane (left panel) and  in the velocity phase plane (right panel).
The solid dot stands for the initial position. For the numerical values of the parameters see main text.}
\label{fig:Fxy-1}
%\end{center}
\end{figure}
\begin{figure}[t]
\centering
%\begin{center}
\includegraphics[scale=0.8]{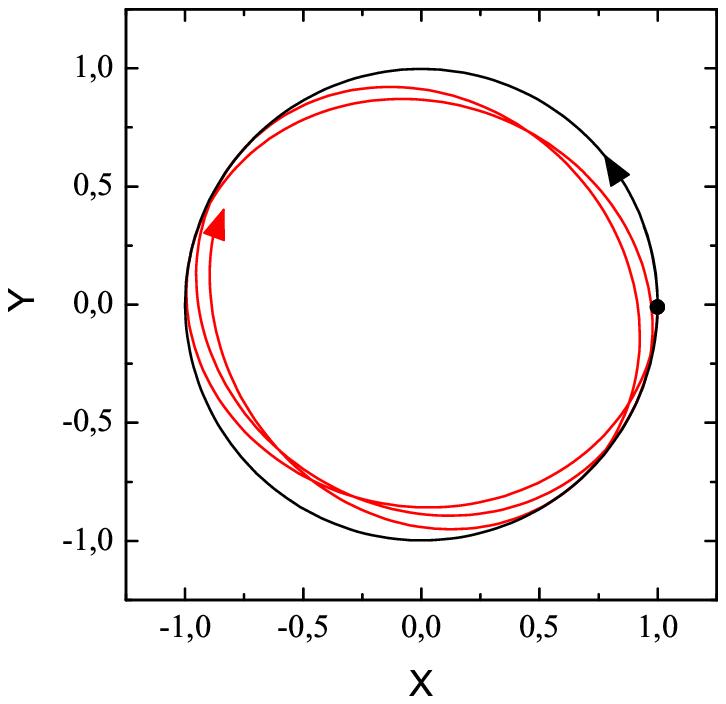}
\includegraphics[scale=0.8]{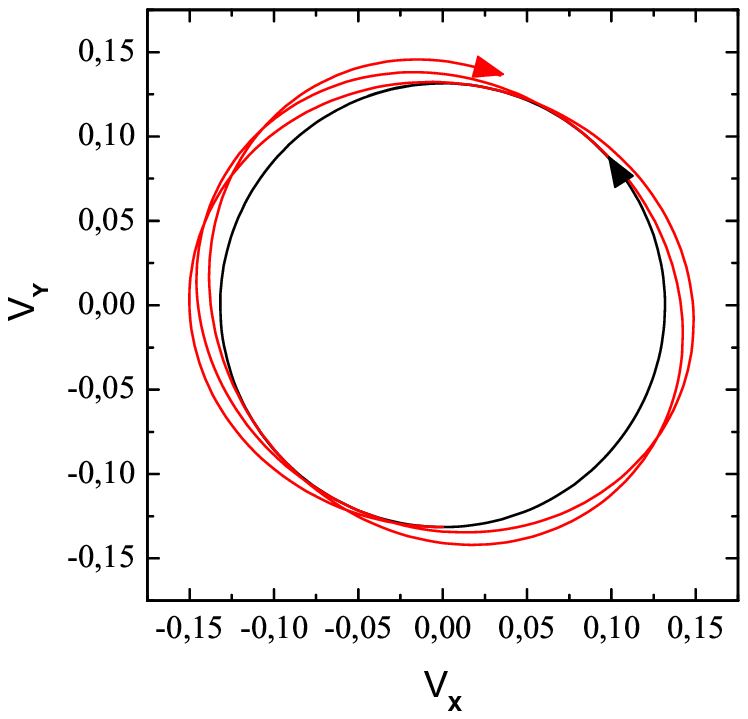}
\caption{Bravais pendulum in two dimensions. Trajectories in phase plane (left panel) and  in the velocity phase plane (right panel).
The black solid line represents the trajectory when the bob turns in the same direction as $\vec \Omega$. The red dotted line corresponds to a motion started with the reversed initial velocity.
The solid dot stands for the initial position. For the numerical values of the parameters see main text.}
\label{fig:Bxy-1}
%\end{center}
\end{figure}
\begin{figure}[t]
\centering
%\begin{center}
\includegraphics[scale=0.8]{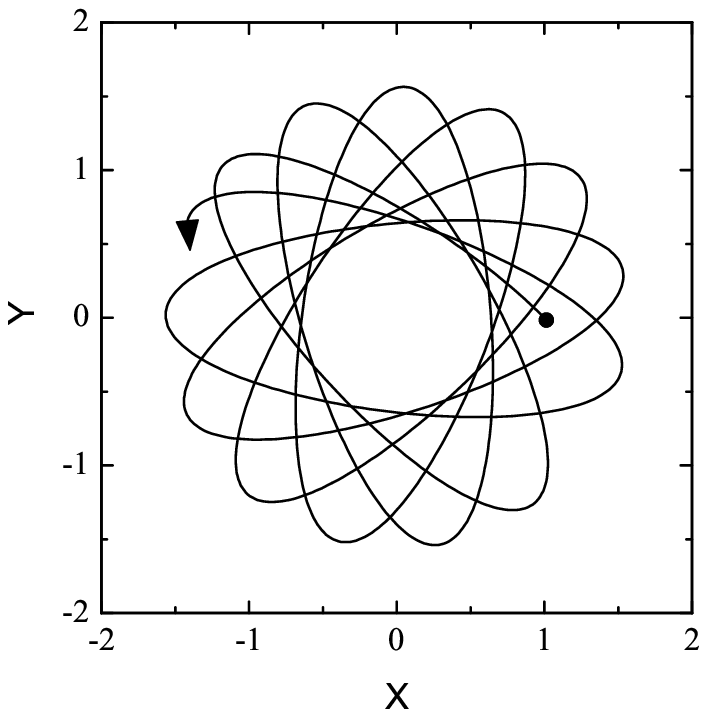}
\includegraphics[scale=0.8]{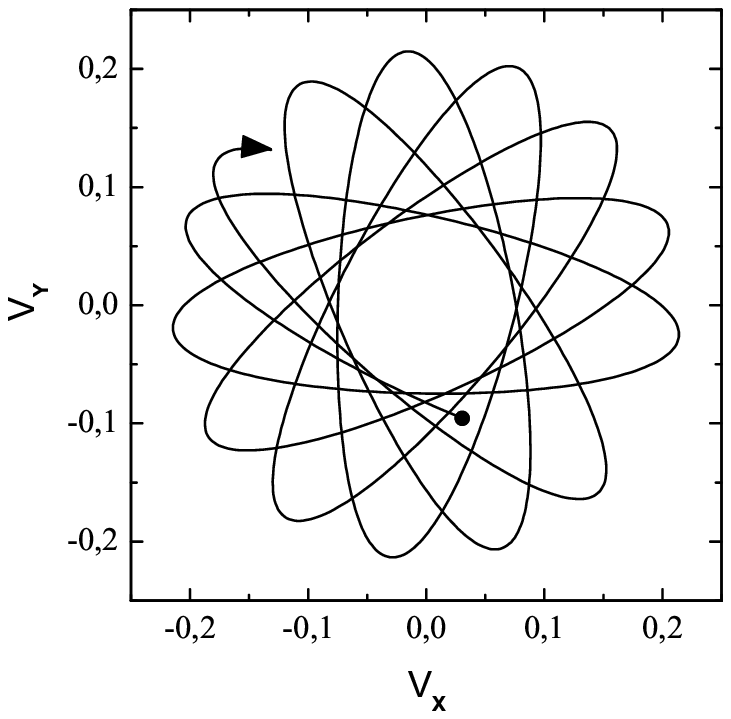}
\caption{General pendulum in two dimensions. Trajectory in phase plane (left panel) and  in the velocity phase plane (right panel).
The solid dot stands for the initial position. For the numerical values of the parameters see main text.}
\label{fig:Gxy-1}
%\end{center}
\end{figure}

\section{The linear pendulum in three dimensions: Exact solution} \label{sec:exact}
Let us consider the equations of motion (\ref{eq:Newtonrotfin}) of the pendulum in three
dimensions without any further approximation, namely, we keep both Coriolis and centrifugal accelerations
\begin{equation}\label{eq:F3vec}
    \frac{d^2 \vec{r}}{dt^2} +2\vec{\Omega} \times \frac{d \vec{r}}{dt}
    + \vec{\Omega} \times (\vec{\Omega} \times \vec{r})+\omega^2\vec{r}=0 .
\end{equation}
The vector products above, being linear operations,  admit a matrix representation. For instance
\begin{equation}\label{eq:A-vec}
    \vec \Omega \times \vec r = \Omega \left(
                                  \begin{array}{ccc}
                                  0 & -\sin \beta & 0 \\
                                 \sin\beta & 0 & \cos\beta \\
                            0&-\cos\beta & 0 \\
                              \end{array}
                          \right) \vec r \equiv  \Omega \Lambda \vec r .
\end{equation}
Equation (\ref{eq:A-vec}) defines the matrix $\Lambda$ of the cross
product associated to the unit vector $\hat \Omega$.
By the same token, we have  $\vec \Omega \times \dot{\vec r}\equiv \Omega \Lambda \dot{\vec r}$,
and  $\vec \Omega \times(\vec \Omega \times {\vec r})\equiv \Omega^2\Lambda^2{\vec r}$.
Now (\ref{eq:F3vec}) may be written down in matrix form
\begin{equation}\label{eq:F3mat}
    \frac{d^2 \vec{r}}{dt^2} +2\Omega \Lambda \frac{d \vec{r}}{dt}
    + [\omega^2 +\Omega^2 \Lambda^2 ] \vec{r}=0 ,
\end{equation}
which we solve next.

We commence with a common exercise posed to undergraduate students in courses of
ordinary differential equations. They are asked to transform
the scalar second order differential equation
\begin{equation}\label{eq:EDO1}
    \frac{d^2 y}{dx^2}+a(x)\frac{dy}{dx}+b(x)y=0,
\end{equation}
into its canonical form, namely the one where the coefficient of the first order derivative vanishes.
To this end one proposes a factorization of the dependent variable
$y(x)$ as $y(x)=u(x)v(x)$, with $v(x)$ the new dependent variable and $u(x)$ to be determined later on.
Substitution in (\ref{eq:EDO1}) yields
\begin{equation}\label{eq:EDO3}
    \frac{d^2 v}{dx^2}+\left( \frac{2}{u}\frac{du}{dx}+a\right) \frac{dv}{dx} +
    \frac{1}{u}\left(  \frac{d^2 u}{dx^2} + a \frac{d y}{dx}+b u\right)=0 .
\end{equation}
If we chose
\begin{equation}\label{eq:u}
    u(x)=\exp\left[ -\frac{1}{2}\int a(s) ds\right],
\end{equation}
then we eventually get the canonical form
\begin{equation}\label{eq:EDO4}
    \frac{d^2 v}{dx^2}+\left[ b(x)-\frac{1}{4}(a(x))^2-\frac{1}{2}\frac{d}{dx}a(x)\right] v=0 .
\end{equation}

To put forward now the question of finding the canonical form of matrix equation
(\ref{eq:F3mat}) is in order. Following a similar procedure as above, we propose the substitution
\begin{equation}\label{eq:change}
    \vec{r}(t)\equiv U(t) \vec{q}(t) ,
\end{equation}
in (\ref{eq:F3mat}), where $U$ is a square real matrix function of dimension three.
Then,   the canonical matrix differential equation, analogous to (\ref{eq:EDO3}), reads
\begin{equation}\label{eq:U}
    \frac{d}{dt}U+\Omega \Lambda U=0,
\end{equation}
Its solution reads
\begin{equation}\label{eq:Uexp}
    U(t)=C \exp(-\Omega \Lambda t), \qquad  C\in \mathcal{R}^{3\times 3}.
\end{equation}
Since the particular value of the constant matrix $C$ is irrelevant for our purposes, we choose the simplest
option, namely $C=I$. As $U(t)$ is not singular we get, after some algebra, the canonical form
\begin{equation}\label{eq:canon}
    \frac{d^2\vec q}{dt^2}+\omega^2 \vec q=0.
\end{equation}
Its solution in terms of initial conditions reads
\begin{equation}\label{eq:q}
    \vec{q}(t)=\cos (\omega t) \, \vec{q}(0)+\frac{1}{\omega}\sin(\omega t)\, \dot{\vec{q}}(0).
\end{equation}
Thus, it turns out that the idea of finding the canonical form of the equations of motion has led to an uncoupled problem.
Moreover, equation (\ref{eq:canon}) has an important dynamical interpretation. It tells us that $U(t)$ transforms the dynamics into three free uncoupled oscillators. We know that this picture corresponds to an analysis in the inertial coordinate system fixed in space (i.e., not turning with the Earth). In this reference system the motion of the pendulum is observed as the composition of three free harmonic oscillators along the axes $x,y,z$.

At this point a comment  about the rhs of (\ref{eq:Uexp}) is in order. The exponential of a square matrix
$B$ is defined by the Taylor expansion:
\begin{equation}\label{eq:B}
    \exp B= I+B+\frac{1}{2!}B^2+\frac{1}{3!}B^3 \ldots
\end{equation}
In general, the problem of finding the exponential of a matrix may be involved, but efficient methods do exist \cite{VanLoan,Gantmacher}. However, due to the symmetry of the matrix $\Lambda$, in our case the computation may be easily carried out by inspection of the powers $\Lambda^n$. It is easy to verify that
\begin{equation}\label{eq:An}
    \Lambda^3= -\Lambda , \qquad \Lambda^4=- \Lambda^2.
\end{equation}
So, recurrent substitution in the Taylor expansion yields
\begin{equation}\label{eq:exp}
    \exp (\Omega \Lambda t)= I-\sin(\Omega t) \Lambda+(1-\cos \Omega t)\Lambda^2 ,
\end{equation}
which provides us with an explicit matrix representation of $U(t)$.

To recover the solution in the rotating reference system fixed to the Earth we use  equation (\ref{eq:change})
\begin{equation}\label{eq:r}
    \vec{r}(t)=U(t)\vec{q}(t)=\mathrm{e}^{-\Omega \Lambda t}\left[ \cos (\omega t) \, \vec{q}(0)+\frac{1}{\omega}\sin(\omega t)\, \dot{\vec{q}}(0)\right] .
\end{equation}
We still have to establish the relationship between the initial conditions in the inertial reference system:
${\vec{q}}(0),\dot{\vec{q}}(0)$;  and those in the rotating reference system fixed to the Earth: ${\vec{r}}(0),\dot{\vec{r}}(0)$. Equation (\ref{eq:r}) and its derivative at time $t=0$ give
\begin{equation}\label{eq:IC}
    {\vec{r}}(0)= {\vec{q}}(0), \qquad  \dot{\vec{r}}(0)=\dot{\vec{q}}(0)+\Omega \Lambda {\vec{q}}(0).
\end{equation}
The exact solution in terms of initial conditions, in the system of reference that rotates with the Earth,
after some algebra is found to be
\begin{eqnarray}\label{eq:F3-fin-matrix} \fl
   \vec{r}(t) = \frac{1}{\omega} \{  [ \omega (1-\cos \Omega t)\cos\omega t -\Omega\sin \omega t \sin \Omega t ]\,\Lambda^2 \, \vec{r}(0)
   +(1-\cos \Omega t)\sin \omega t \,\Lambda^2 \, \dot{\vec{r}}(0)  \nonumber\\ \fl
   \phantom{\vec{r}(t) =}
   +[\Omega \cos \Omega t \sin \omega t -\omega \sin \Omega t \cos \omega t ]\, \Lambda \,{\vec{r}}(0) -
   \sin \Omega t \sin \omega t \, \Lambda \, \dot{\vec{r}}(0)   \nonumber \\  \fl
   \phantom{\vec{r}(t) =}  +\omega \cos \omega t \, {\vec{r}}(0) +\sin \omega t \, \dot{\vec{r}}(0) \} .
\end{eqnarray}
Taking into account trigonometric identities for the product of $\sin$ and $\cos$ functions we can conclude that the motion is given by the superposition of three frequencies: $\omega \pm \Omega$, and $\omega$. This is at variance with the approximate trajectories in two dimensions (\ref{eq:traj-x}) and (\ref{eq:traj-y}) where only the two frequencies: $\xi\pm \tilde \Omega \simeq \omega \pm s\Omega$, are present. In the case of the Earth rotation, it must not be relevant, for we should be able to resolve between two and three peaks in the power spectrum of, say, $x(t)$; where $\Omega \ll \omega$.

Since that $\Lambda$ is the matrix representation of the vector product associated
to the unit vector  $\hat \Omega$, we can also write
\begin{eqnarray}\label{eq:F3-fin-vec} \fl
  \vec{r}(t) =\frac{1}{\omega }\{ [ \omega (1-\cos \Omega t)\cos\omega t -\Omega\sin
  \omega t \sin \Omega t ]\;\hat \Omega \times( \hat \Omega \times \vec{r}(0))
  \nonumber \\   \fl
  \phantom{\vec{r}(t) =}+[(1-\cos \Omega t)\sin \omega t ]\;
  \hat \Omega \times (\hat \Omega \times \dot{\vec{r}}(0))
  \nonumber \\   \fl
  \phantom{\vec{r}(t) =}
    +[\Omega \cos \Omega t   \sin \omega t -\omega \sin \Omega t \cos \omega t ]\; \hat \Omega \times{\vec{r}}(0)
   \nonumber \\ \fl
   \phantom{\vec{r}(t) =} -[\sin \Omega t \sin \omega t ] \; \hat \Omega \times \dot{\vec{r}}(0)
    +\omega \cos \omega t \; {\vec{r}}(0) +\sin \omega t \; \dot{\vec{r}}(0) \}  .
\end{eqnarray}
Notice that in the limit $\Omega \to 0$, the equation above gives the expected result
$\vec{r}(t) =\cos \omega t \; {\vec{r}}(0) +\frac{1}{\omega}\sin \omega t \; \dot{\vec{r}}(0) $.

In terms of components the exact solution reads
\begin{eqnarray} \label{eq:F3-fin-compx} \fl
  x(t) = \left\{  [ c^2+\cos \left( \Omega\,t \right) s^2   ] \,\cos \left(\omega\,t \right)
+\frac{\Omega}{\omega}\sin(\Omega \, t)\sin(\omega \, t) \, s^2  \right\}
{ x(0)}\nonumber \\ \fl \phantom{x(t) =}
-  \left[ \frac{\Omega}{\omega}  \cos \left( \Omega\,t \right) \sin \left( \omega\,t
 \right) -
\sin \left( \Omega\,t \right) \cos \left( \omega\,t \right)
  \right]  \, s \,{ y(0)}\nonumber \\ \fl \phantom{x(t) =}
+\left\{ \frac{\Omega}{\omega}\sin(\Omega \, t)\sin(\omega \, t)
- [ 1-\cos \left( \Omega\,t \right)  ]  \,\cos \left( \omega\,t \right) \right\} \, sc \,
{ z(0)}   \nonumber \\ \fl \phantom{x(t) =}
 + \frac{ \sin  \omega\,t   }{\omega} \Bigl\{ [ c^2+
 \cos \left( \Omega\,t \right)    s^2  ] {
 \dot x (0)}
+\sin \left( \Omega\,t \right)
s\, { \dot y (0)} - \left[ 1-\cos \left( \Omega\,t
 \right)  \right]\, sc\,
 { \dot z (0)} \Bigr\}
 \\ \nonumber
 \\ \label{eq:F3-fin-compy} \fl
  y(t) =  \left[ \frac{\Omega}{\omega} \cos \left( \Omega\,t \right) \sin \left( \omega\,t
 \right)
  -\sin \left( \Omega\,t \right) \cos \left( \omega\,t
 \right)
 \right] [ s \, { x(0)} + c\, z(0) ]
   \nonumber \\ \fl \phantom{y(t) =}
 +\left\{ \frac{\Omega}{\omega} \sin(\Omega \, t) \sin(\omega \, t)
 +\cos \left( \Omega\,t \right) \cos \left(
\omega\,t \right) \right\} { y(0)}
 \nonumber \\ \fl \phantom{y(t) =}
- \frac{\sin  \omega\,t}{\omega}  \Bigl[ \sin \left( \Omega\,t
 \right)  s \, {
\dot x (0)}-\cos \left( \Omega\,t \right)  {
\dot y (0)}
+\sin \left( \Omega\,t \right) c\, { \dot z (0)}\Bigr]
\\ \nonumber
\\ \label{eq:F3-fin-compz}\fl
  z(t)=  \left\{  \frac{\Omega}{\omega} \sin(\Omega \, t) \sin(\omega \, t)
  -\left[ 1-\cos \left( \Omega\,t \right)  \right] \, \cos \left( \omega\,t
 \right) \right\} \, sc \,
 { x(0)}\nonumber \\ \fl \phantom{z(t) =}
 + \left[\sin \left( \Omega\,t \right) \cos \left( \omega\,t \right) -\frac{\Omega}{\omega}\cos
 \left( \Omega\,t \right) \sin \left( \omega\,t \right) \right] c\,  { y(0)}\nonumber \\ \fl \phantom{z(t) =}
+ \left\{ \left[ s^2+ \cos \left(
\Omega\,t \right)   c^2
 \right] \cos \left( \omega\,t \right) + \frac{\Omega}{\omega} \sin(\Omega \, t) \sin(\omega \, t) \, c^2 \right\}
 { z(0)}
  \nonumber \\ \fl \phantom{z(t) =}
 - \frac{\sin  \omega\,t }{\omega}  \Bigl\{ \left[ 1
-\cos \left( \Omega\,t \right)  \right] sc\,  { \dot x (0)}
 -\sin
 \left( \Omega\,t \right)  c\, { \dot y (0)}
- \left[ s^2+ \cos \left( \Omega\,t
 \right)    c^2
 \right]  { \dot z (0)} \Bigr\} ,
\end{eqnarray}
where $s,c$ have been defined in (\ref{eq:sc}).

We have used the expressions above to plot the  trajectories corresponding to the very same initial conditions as in figures of Section \ref{sec:2D}. The paths are rather different.

\subsection{Foucault pendulum in 3D}
\begin{figure}[t]
\centering
%\begin{center}
\includegraphics[scale=0.8]{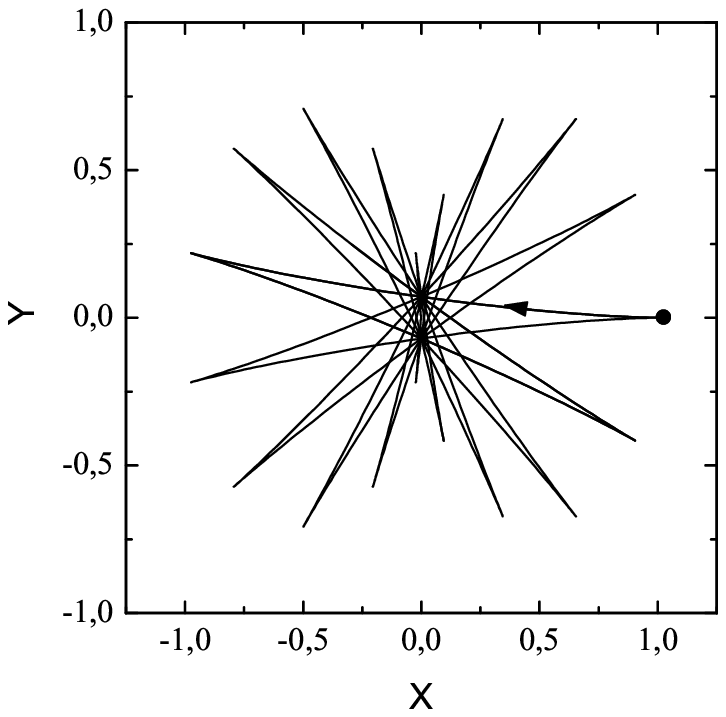}
\includegraphics[scale=0.8]{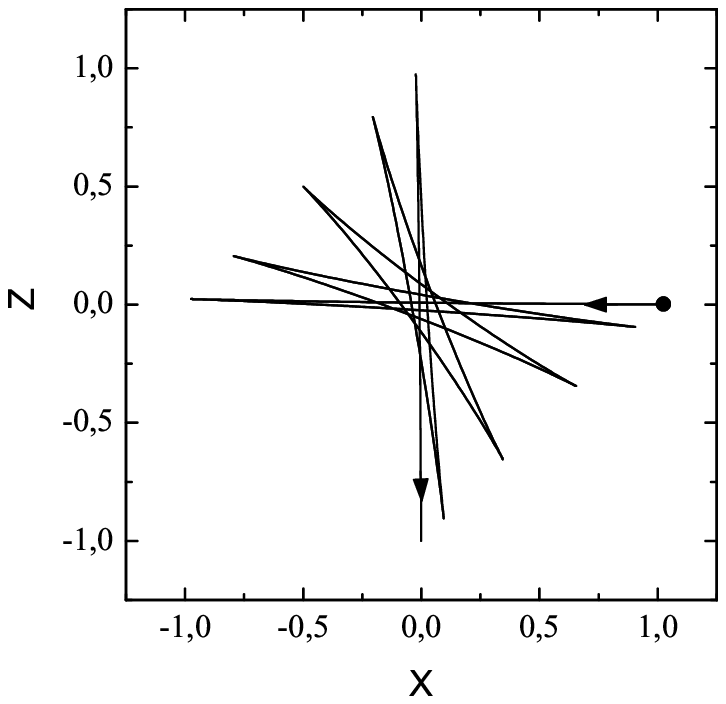}
\caption{Foucault pendulum in three dimensions. Trajectory in the phase plane $xy$ (left panel) and  in the the phase plane $xz$ (right panel).
The solid dot stands for the initial position. For the numerical values of the parameters see main text.}
\label{fig:Fxy}
%\end{center}
\end{figure}

In Figure \ref{fig:Fxy} we represent the exact trajectory corresponding to the Foucault pendulum with the same initial conditions as in Figure \ref{fig:Fxy-1}. The left panel presents the $xy$ projection and can be compared with the pattern in Figure \ref{fig:Fxy-1} (left panel). Notice the difference in symmetry. The right panel gives the $xz$
projection. In the two--dimensional case it is simply $z(t)=0$.

\subsection{Bravais pendulum in 3D}
\begin{figure}[t]
\centering
%\begin{center}
\includegraphics[scale=0.8]{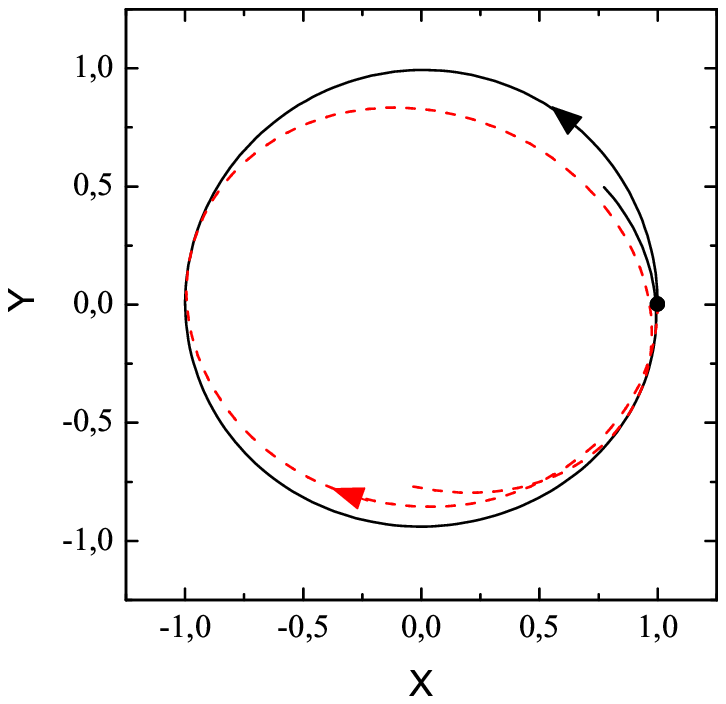}
\includegraphics[scale=0.8]{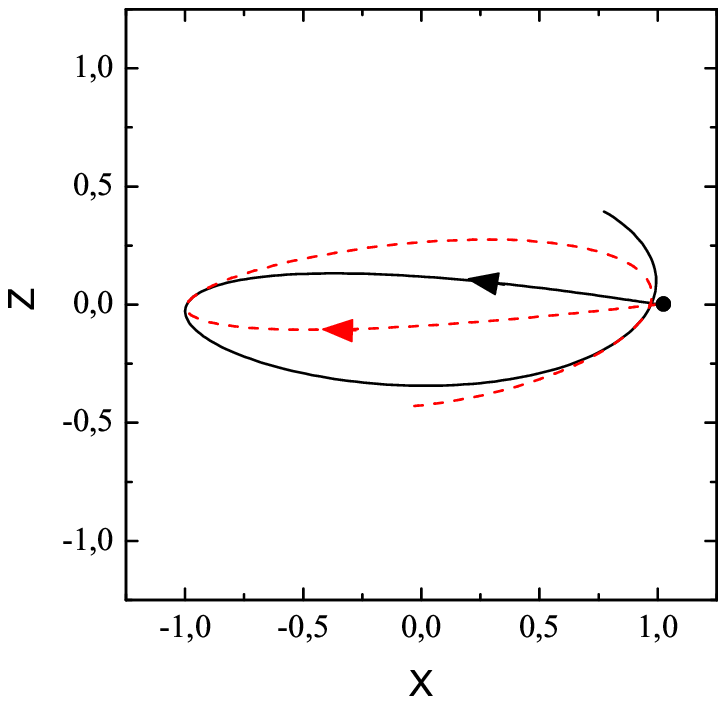}
\caption{Bravais pendulum in three dimensions. Trajectories in the phase plane $xy$ (left panel) and  in the  the phase plane $xz$ (right panel).
The black solid line represents the trajectory when the bob turns in the same direction as $\vec \Omega$. The dotted line
corresponds to a motion started with the reversed initial velocity.
The solid dot stands for the initial position. For the numerical values of the parameters see main text.}
\label{fig:Bxy}
%\end{center}
\end{figure}

The Bravais pendulum trajectories are represented in Figure \ref{fig:Bxy}. The solid black line corresponds to a turn parallel to $\vec \Omega$, and the dotted red line is the trajectory with the initial tangential velocity reversed.
We have used the same condition for the initial velocity $\dot y(0)$ as in the planar case. It turns out that after
one cycle the oscillation is not conical-like any longer. The time delay between both turning modes is also apparent, as in the two-dimensions case.

\subsection{General pendulum in 3D}
\begin{figure}[t]
\centering
%\begin{center}
\includegraphics[scale=0.8]{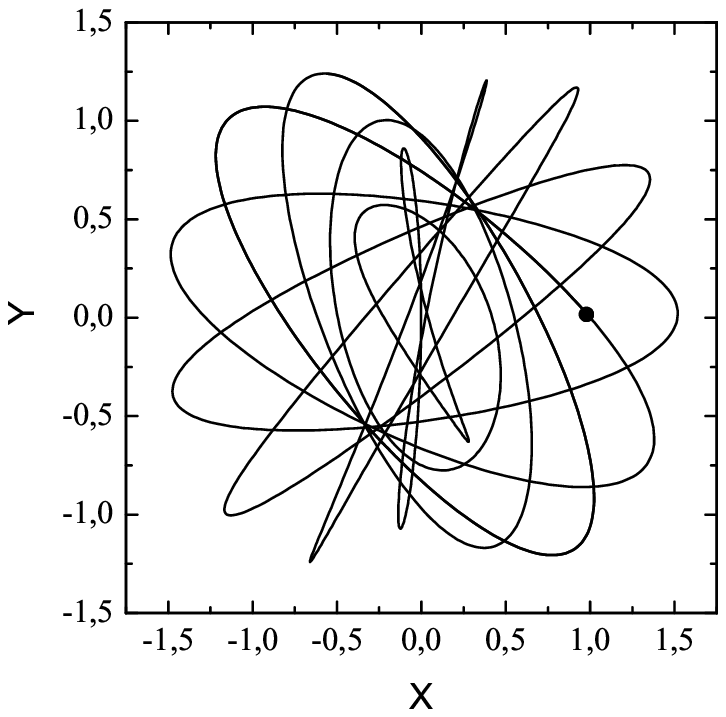}
\includegraphics[scale=0.8]{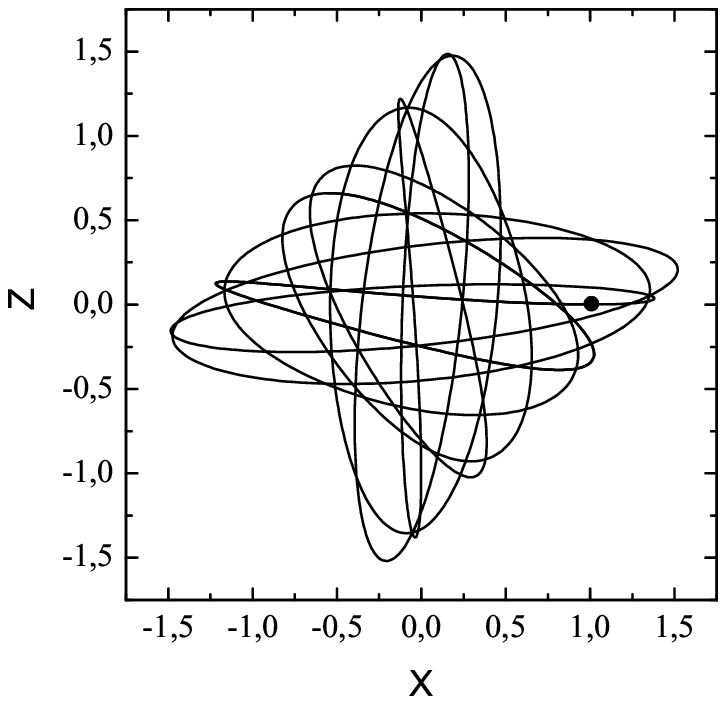}
\caption{General pendulum in three dimensions. Trajectory in the phase plane $xy$ (left panel) and  in the  the phase plane  $xz$ (right panel).
The solid dot stands for the initial position. For the numerical values of the parameters see main text.}
\label{fig:Gxy}
%\end{center}
\end{figure}

The more general mode of oscillation of Figure \ref{fig:Gxy-1} is replicated in Figure \ref{fig:Gxy} for the three--dimensional pendulum. The left and right panels give the $xy$ and $xz$ projections respectively. As in the two preceding subsections, the differences in the patterns are noteworthy.
We will discuss about these differences in the final Section.


\section{Chevilliet's theorem} \label{sec:Chevilliet}
Let us consider the case where we release the pendulum bob at
$\vec{r}(0)=(\epsilon ,0,\lambda)^{\top}$, from rest: $\dot{\vec{r}}(0)=0$, with $\lambda \equiv l-\sqrt{l^2-\epsilon ^2}$, and $\epsilon \ll l$.
It is then interesting to study the trajectory as it is seen from the inertial reference system fixed in space.
According to (\ref{eq:IC}), the corresponding initial conditions are
$\vec{q}(0)=(\epsilon ,0,\lambda)^{\top}, \dot{\vec{q}}(0)=(0,-\Omega[s\epsilon +c\lambda] ,0)^{\top}$,
where $s,c$ have been defined in (\ref{eq:sc}). Then (\ref{eq:q}) gives
\begin{eqnarray} \label{eq:F3-s-vec-0}
  q_x(t) &=& \epsilon \,\cos(\omega t),\nonumber \\
  q_y(t) &=& -\frac{\Omega}{\omega}[s\epsilon+c\lambda]\, \sin(\omega t) , \\
  q_z(t) &=& \lambda \, \cos(\omega t) . \nonumber
\end{eqnarray}
It is straightforward to see that, in three dimensions, the observed trajectory takes place on the ellipsoid
\begin{equation}\label{eq:F3-elipsoide}
    \frac{q^2_x+q^2_z}{\epsilon^2+\lambda^2}+  \frac{q^2_y}{(s\epsilon+c\lambda)^2}\cdot \frac{\omega^2}{\Omega^2}=1.
\end{equation}
The intersection with the plane $q_z=0$ is the ellipse
\begin{equation}\label{eq:F3-elipse2D}
    \frac{q^2_x}{\epsilon^2+\lambda^2}+ \frac{q^2_y}{(s\epsilon+c\lambda)^2}\cdot \frac{\omega^2}{\Omega^2}=1,
\end{equation}
whose ratio of axes in the  limit $\epsilon \ll l$ can be approximated by
\begin{equation}\label{eq:F3-semiejes}
    \frac{\Omega}{\omega}\left( s+c\frac{\epsilon}{2l}  \right),
\end{equation}
since $\lambda \simeq \epsilon^2/2l$. Thus, the planar projection of the pendulum motion is an ellipse as a consequence
of the non--vanishing initial (tangential) velocity provided by the rotation of the Earth on the
bob. Otherwise the trajectory would be on a straight segment.

Had we resolved the same initial value problem with the approximated  pendulum equations in two dimensions of
Section \ref{sec:2D}, we would have found that the trajectory in the fixed reference system
describes the ellipse
\begin{equation}\label{eq:Fouc-asim-ellipse0}
    \frac{q^2_x}{\epsilon ^2}+\frac{q^2_y}{\epsilon ^2}\cdot \left(\frac{\omega}{s\Omega}\right)^2=1 ,
\end{equation}
whose axes ratio, $s\Omega/\omega$,  is determined simply by the ratio of the Earth rotation frequency
at latitude  $\beta$, to that of the natural frequency of the pendulum.
This is nowadays an almost forgotten result, known in old texts as
\emph{Chevilliet theorem} \cite{Kimball, MacMillan, Appell}. Interestingly, we can also recover this theorem
by doing a further approximation in (\ref{eq:F3-elipse2D}). Indeed, the drastic approximation $\lambda=0$
immediately leads to the Chevilliet theorem. Thus, the solution in three dimensions (\ref{eq:F3-semiejes})
provides a small correction to Chevilliet's ellipse (\ref{eq:Fouc-asim-ellipse0}).
We are now ready to accept that the convoluted patterns we observe in figures such as
Figure \ref{fig:Fxy-1} is the direct consequence of observing the motion on a Chevilliet ellipse from a rotating reference system.

We can buttress the idea of the \emph{geometric} origin of the convoluted patterns in the phase plane as follows.
Consider, in the exact solution in (\ref{eq:F3-fin-vec}), the subtraction
\begin{equation}\label{eq:sustrae}
\vec{S}(t)\equiv \vec{r}(t) -[\omega \cos \omega t \; {\vec{r}}(0) +\frac{1}{\omega}\sin \omega t \; \dot{\vec{r}}(0) ].
\end{equation}
We can interpret $\vec S$ as the remaining line of motion that is left once the elliptic piece has been removed (i.e, the non--perturbed or inertial part) from the exact trajectory.
Next, due to the mathematical properties of the cross and scalar products, we get for that remainder: $\hat \Omega \cdot \vec{S}(t)=0,\forall  t$. Hence, every additional contribution to the trajectory originated in Coriolis and centrifugal acceleration terms takes place in a plane orthogonal  to $\hat \Omega$. There is no correction to the trajectory in the very direction of the Earth angular velocity vector $\vec \Omega$.

The flowery nature of the patterns of the pendulum trajectories emerges, therefore, from the non--inertial character of the rotating frame. Everything should appear as an ellipse in the phase plane if we were able to choose the proper reference system, so to speak.

\begin{figure}[t]
\centering
%\begin{center}
\includegraphics[scale=0.8]{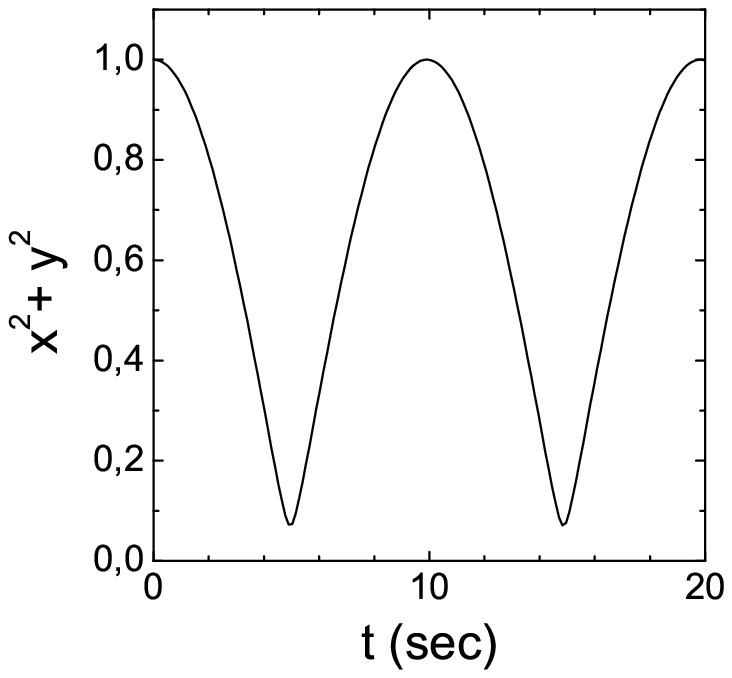}
\includegraphics[scale=0.8]{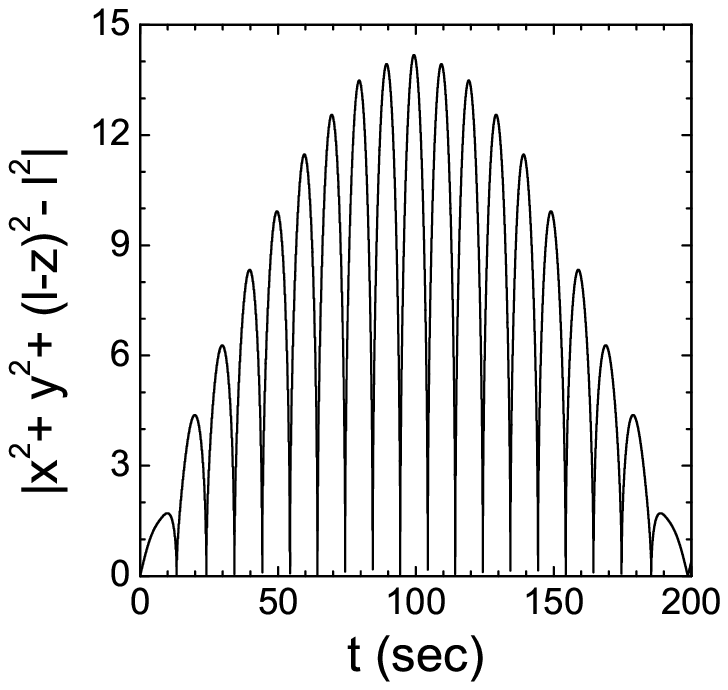}
\caption{Foucault pendulum in two (left panel) and three (right panel) dimensions.
The functions exhibit the non--preservation of the constancy of the pendulum length $l$ (preservation corresponds to
the value zero in the plots).}
\label{fig:L}
%\end{center}
\end{figure}

Finally, neither the solution in two nor the one in three dimensions preserves the constancy of the length of the pendulum. This is not a consequence of the non-vanishing angular velocity of the reference system but the approximations made to get a linear systems of differential equations. It is a price to pay once we have replaced a non--linear problem (the physical pendulum) with a linear one (the simple pendulum).
Namely, the constraint $x^2+y^2+(l-z)^2= l^2$ (or, equivalently, $q_x^2+q_y^2+(l-q_z)^2= l^2$) which represents a sphere
of radius $l$, centered in $(0,0,l)$, is not preserved along a trajectory. Instead of a sphere we have  the ellipsoid (\ref{eq:F3-elipsoide}). The same drawback occurs in the approximate treatment in two dimensions. There the constraint is $x^2+y^2=0$ (or, $q_x^2+q_y^2=0$), which is by no means preserved but replaced with the  Chevilliet ellipse constraint (\ref{eq:Fouc-asim-ellipse0}).   Figure \ref{fig:L} illustrates this discrepancy in two (left panel) and three (right panel) dimensions as a function of time. Both functions are periodic but with different periods.
For a pendulum length $l=100$ m., in the two dimension model the period is $20$ sec., and the maximal relative deviation is $1/100$. In three dimensions the period is $200$ sec., and the the maximal relative deviation is $14/100$.

\section{Foucault pendulum: Comparison of the  2D and 3D cases }\label{sec:2D3D}

At this point we have already realized that the trajectories obtained from the 2D--  and the $xy$ phase plane from the 3D--system of equations do differ. Next we study whether these differences are relevant for the analysis of Foucault pendula placed on Earth.

In his original papers \cite{travaux} Foucault explains how in Paris the pendulum takes \emph{circa} 30 hours to complete one turn, which corresponds to one degree in five minutes. This is to say that the museum common visitor observes, at most, that the pendulum bob describes a small arc of circumference during her or his visit. In the original setup at the \emph{Panth\'eon} there was a pile of damp sand disposed in circular shape around the pendulum equilibrium point that a stylus attached to the underside of the bob plowed, witnessing the accumulative deflection of the trajectory. Modern museums setups rather use falling rods or balls instead of sand.

\begin{figure}[t]
\centering
%\begin{center}
\includegraphics[scale=0.8]{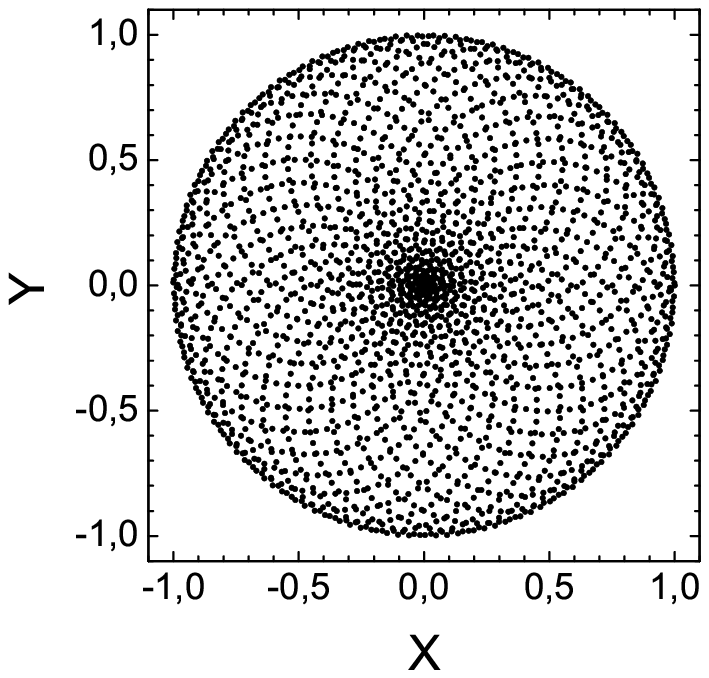}
\includegraphics[scale=0.8]{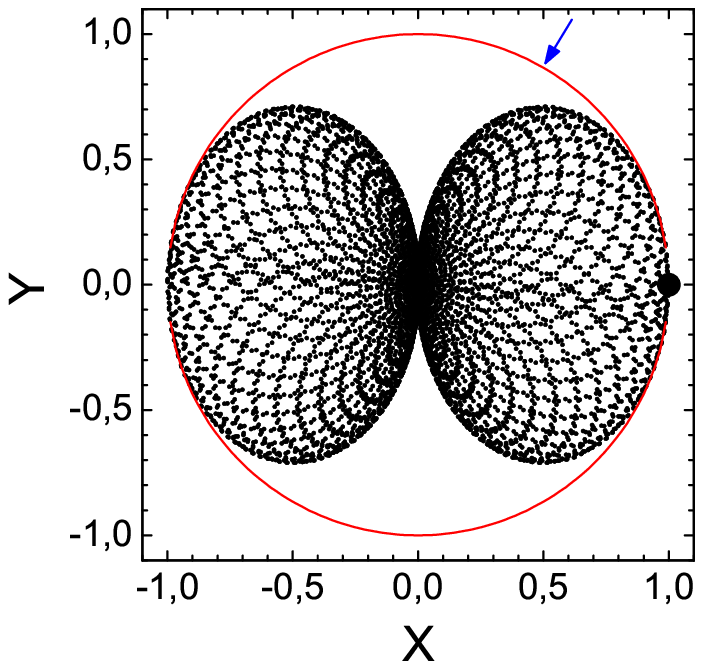}
\caption{Foucault pendulum. Black dots represent the area swept by the bob after one complete turn in the two (left panel) and three (right panel) dimensional model. The red solid line corresponds to the border of the pattern in the left panel and is given for the sake of comparison. The blue arrow locates $60^0$ deflection of the oscillation plane with respect the initial position (large solid black dot).}
\label{fig:Earth}
%\end{center}
\end{figure}

The solutions (\ref{eq:traj-x}) and (\ref{eq:traj-y}) of the 2D model show that the bob sweeps a circular area after a complete turn. This is represented in the left panel of Figure \ref{fig:Earth}. The black dots stand for the location of the bob, sampled every $0.4$ sec. (no need to seek any especial pattern inside). The pendulum length is $100$ m., and $\Omega=0.000073$ rd/sec., is the real Earth angular velocity. The bob is released from rest (in the Earth attached reference system) at $x(0)=1$ m., $y(0)=0$. In addition, the right panel in Figure \ref{fig:Earth} represents the corresponding 3D case from (\ref{eq:F3-fin-compx}) and (\ref{eq:F3-fin-compy}). Here the red solid line corresponds to the pattern border in the preceding panel, and it is included for the sake of comparison.
Definitely, this dual cardioid--like shape is not the observed pattern in either museums  or reported by Foucault.

There is an untold piece in this story that reasonably explains the disagreement. Air resistance gradually damps the oscillation of the pendulum. Foucault wrote \cite{travaux} that every five or six hours (\emph{i.e.}, deflection of 60 to 70 degrees) the experiment must be stopped in order to restart it from that very precise position with the original oscillation amplitude restored. The blue arrow in Figure \ref{fig:Earth} points out that event and, without further information, the receding of the black border could be erroneously ascribed to damping. Of course, the air resistance effect is incremental.

The course of the experiment is then punctuated by restarting and, as a consequence,  what one sees is the juxtaposition of several copies of the first $60$ degrees, an (almost) circular sector of the pattern in the right panel of Figure \ref{fig:Earth}. The overall composition mimics the circular pattern.

Thus damping prevents the experimental observation of the
dual cardioid--like pattern. The decreasing of the oscillation amplitude is caused by the constructive combination of damping and very three dimensional nature of the motion.
Indeed, Foucault's experiment was not designed to resolve these two contributions but to exhibit the deflection of the rotation plane of the pendulum. What we argue is that, in practical realizations, both panels in Figure \ref{fig:Earth} are not in conflict.

In 1855 a Foucault pendulum was established as permanent exhibit  at \emph{Ars et M\'etiers} with the occasion of the Universal Exhibition held in Paris. At this time an ingenious electromagnetic device was added so that the experiment needed no restarting. An electromagnet placed on the ground exerted attraction on the bob only when this was approaching the equilibrium point. When it crossed the vertical and the bob was receding the action of the electromagnet stopped. A mechanical synchronization device restarted the current once the bob was anew in the approaching phase.

Modern pendula exhibitions incorporate also a driving apparatus.
For instance, in the Museum of Arts and Sciences in Valencia there is a permanent exhibit. The pendulum is driven electromagnetically through a system located in its upper part. The device actions are controlled by an electronic automata. The question arises then: to what extent is the attractor of the system modified? We will not enter here into this subject and will limit ourselves just to point out that the solvable, linear, homogeneous,  second order system of differential equations with constant coefficients in (\ref{eq:F3}) becomes a different class of equation: non--constant coefficients, non--homogeneous, or non--linear; depending on the type of driving. In general, non--solvable.

\section{Final comments}\label{sec:fin}
The Foucault pendulum as well as its generalizations, are not only beautiful museum experiments and
excellent tools from a pedagogical point of view (as they contain an ideal admixture of
mechanics and mathematics), but also they do have far reaching applications. The Foucault pendulum-like
motion can be viewed as a relative orbital motion under a central force, like the relative motion
of celestial bodies subject to the influence of the same gravitational field, or terminal
orbital rendezvous and as such the subject is still an active topic and has actual relevance.

The full richness of these systems is sometimes disguised when studied as separate entities. The
unified treatment we propose here manifestly shows the crucial role played by initial conditions
in determining the specific features of the system.
Here, as in spontaneous symmetry breaking, where the Lagrangian possesses a symmetry not respected
by the vacuum, the equations of motion are identical in every case, the selection is done by the
initial conditions.

\ack
%%%%%%%%%%%%%%%%%%%%%%%%%%%%%%%%%%%%%%%%%%%%%%%%%%%%%%%%%%%%%%%%%%%%%%
This work has been partially supported by spanish MEC and FEDER (EC) under grant
FPA2011-23596, and Generalitat Valenciana under the grant PROMETEO/2008/004 (G.B.); and
MICINN (Spain) under grant no. AYA2010-22111-C03-02 (J.A.O.).

 \end{document}